\documentclass[aps,pra,twocolumn,amsmath,amssymb,showpacs,superscriptaddress,nobalancelastpage]{revtex4-1}

\usepackage{graphicx}% Include figure files
\usepackage{dcolumn}% Align table columns on decimal point
\usepackage[mathlines]{lineno}
\usepackage{hyperref}
\usepackage{bm}% bold math
\usepackage{epstopdf}
\usepackage{epsfig}
\usepackage{amssymb}
\usepackage{braket}
\usepackage{amsmath}
\usepackage{mathtools}
\usepackage{color}
\usepackage{soul} 
\usepackage{graphics}

\newcommand{\beginsupplement}{%
        \setcounter{table}{0}
        \renewcommand{\thetable}{S\arabic{table}}%
        \setcounter{figure}{0}
        \renewcommand{\thefigure}{S\arabic{figure}}%
     }

\begin{document}
\bibliographystyle{unsrt}
\title{Multi-dimensional entanglement transport through single-mode fibre}
\date{\today}

\author{Jun Liu}
\email[author contributions:]{These authors contributed equally to the work}
\affiliation{Wuhan National Laboratory for Optoelectronics, School of Optical and Electronic Information, Huazhong University of Science and Technology, Wuhan 430074, Hubei, China.}

\author{Isaac Nape}
\email[author contributions:]{These authors contributed equally to the work}
\affiliation{School of Physics, University of the Witwatersrand, Private Bag 3, Wits 2050, South Africa}

\author{Qianke Wang}
\affiliation{Wuhan National Laboratory for Optoelectronics, School of Optical and Electronic Information, Huazhong University of Science and Technology, Wuhan 430074, Hubei, China.}

\author{Adam Vall\'{e}s}
\affiliation{School of Physics, University of the Witwatersrand, Private Bag 3, Wits 2050, South Africa}

\author{Jian Wang}
%\email[Corresponding author: ]{jwang@hust.edu.cn}
\affiliation{Wuhan National Laboratory for Optoelectronics, School of Optical and Electronic Information, Huazhong University of Science and Technology, Wuhan 430074, Hubei, China.}

\author{Andrew~Forbes}
%\email[Corresponding author: ]{andrew.forbes@wits.ac.za}
\affiliation{School of Physics, University of the Witwatersrand, Private Bag 3, Wits 2050, South Africa}

%_____________________________________________________
\begin{abstract}
\noindent The global quantum network requires the distribution of entangled states over long distances, with significant advances already demonstrated using entangled polarisation states, reaching approximately 1200 km in free space and 100 km in optical fibre. Packing more information into each photon requires Hilbert spaces with higher dimensionality, for example, that of spatial modes of light.  However spatial mode entanglement transport requires custom multimode fibre and is limited by decoherence induced mode coupling.  Here we transport multi-dimensional entangled states down conventional single-mode fibre (SMF). We achieve this by entangling the spin-orbit degrees of freedom of a bi-photon pair, passing the polarisation (spin) photon down the SMF while accessing multi-dimensional orbital angular momentum (orbital) subspaces with the other.  We show high fidelity hybrid entanglement preservation down 250 m of SMF across multiple $2\times2$ dimensions, demonstrating quantum key distribution protocols, quantum state tomographies and quantum erasers.  This work offers an alternative approach to spatial mode entanglement transport that facilitates deployment in legacy networks across conventional fibre.  
\end{abstract}
\date{\today}
\maketitle

%_____________________________________________________
\noindent Entanglement is an intriguing aspect of quantum mechanics with well-known quantum paradoxes such as those of Einstein-Podolsky-Rosen (EPR) \cite{PhysRev.47.777}, Hardy \cite{PhysRevLett.71.1665}, and Leggett \cite{Leggett2003}. Yet it is also a valuable resource to be harnessed: entangled particles shared with different distant observers can be used in quantum cryptography to set an unconditional secure key \cite{PhysRevLett.67.661,PhysRevLett.84.4729}, in quantum teleportation to transfer quantum information \cite{PhysRevLett.70.1895,nature390,nature429,nature4291,nature430}, in super-dense coding \cite{PhysRevLett.69.2881,1367-2630-12-7-073042}, in ghost imaging \cite{PhysRevA.52.R3429,PhysRevA.78.061802}, and are also an important part for quantum computation \cite{2058-7058-11-3-31,nature434,nature445}.

In the past few decades, quantum entanglement has been extensively explored for a variety of quantum information protocols. Standard quantum communication protocols exploit polarisation (or ``spin'' angular momentum) encoding with single photon and multipartite states. Up to now, entanglement transport has been verified over distances up to 1200 km via free space (satellite-based distribution) \cite{Yin1140} and 102 km through fibre using polarisation-entangled photons \cite{Hubel:07}. %However, two-dimensional polarisation entanglement impose significant limitations on the information capacity of the protocols that may be implemented, i.e., only 1 bit per photon can be encoded.
%%%%%%%%%%%%%%%%%%%%%%%%%%%%%%%%%%%%%%
\begin{figure}[h!]
\centering
\includegraphics[width=\linewidth]{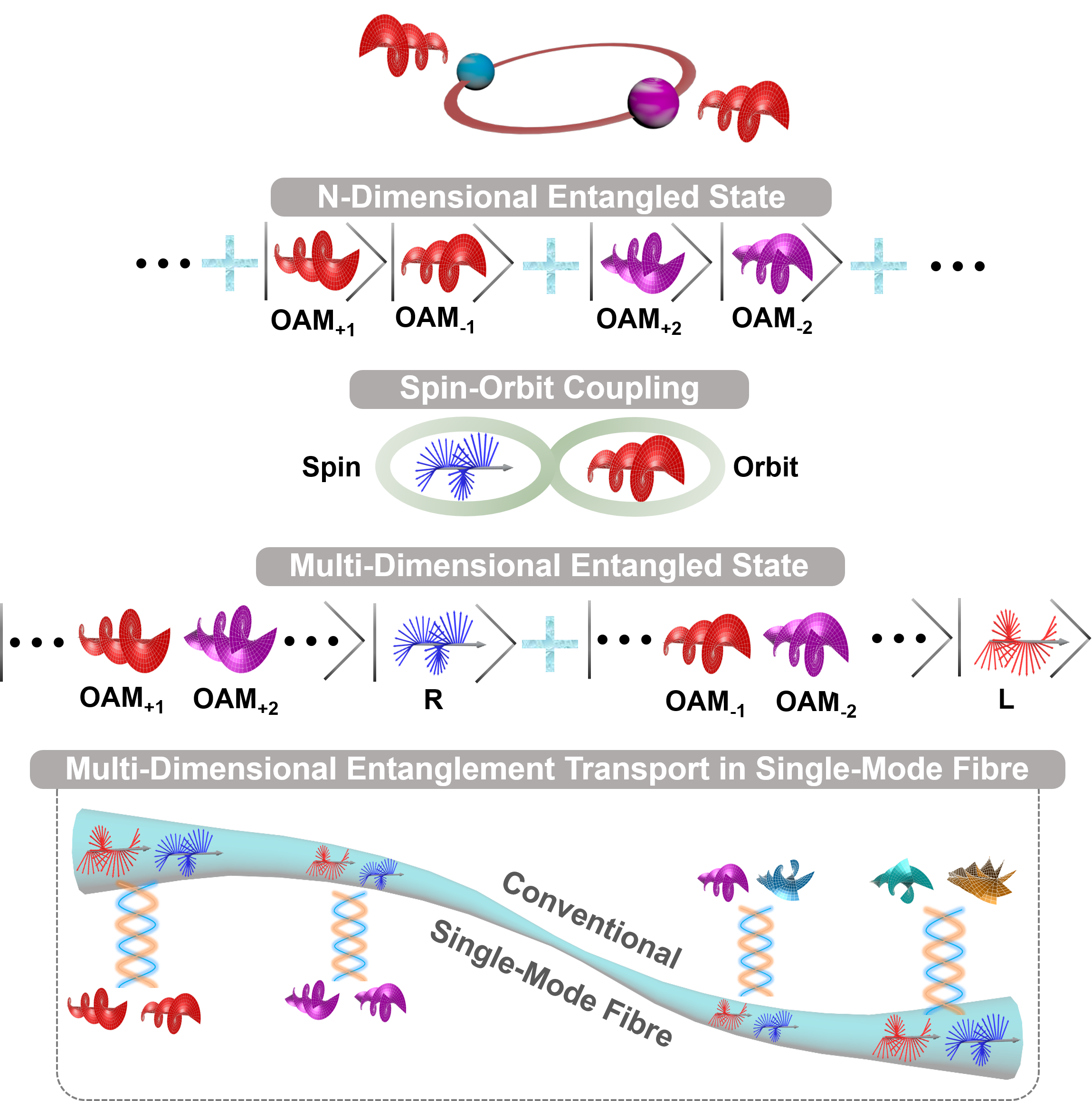}
\caption{\textbf{Multi-dimensional entanglement transport through single-mode fibre.}  An initially high-dimensional state is converted by spin-orbit coupling into multiple two dimensional hybrid entangled states.  The polarisation (spin) photon has a fundamental spatial mode (Gaussian), facilitating transport down a SMF.  By the hybrid entanglement the infinite space of OAM is still available.}
\label{conceptfigure}
\end{figure} 
%%%%%%%%%%%%%%%%%%%%%%%%%%%%%%%%%%%%%%
Exploiting high-dimensional entangled systems presents many opportunities \cite{LSA7}, for example, a larger alphabet for higher photon information capacity and better robustness to background noise and hacking attacks \cite{PhysRevLett.88.127902}. Remarkably, temporal \cite{PhysRevLett.62.2205,PhysRevLett.93.010503}, frequency \cite{PhysRevLett.103.253601}, and spatial  \cite{PhysRevLett.64.2495,nature412,PhysRevLett.103.083602} entanglement can be tailored with photons.  More recently, the orbital angular momentum (OAM) of light, related to the photon's transverse mode spatial structure, has been recognized as a promising resource exploiting high-dimensional states encoded in a single photon \cite{naturephysics3,Krenn20150442}. Quantum communication with spatial modes is still in its infancy, with reported entanglement transport in multi-mode fibre limited to less than 1 m \cite{PhysRevLett.106.240505,PhysRevLett.109.020502}.  While new (under review) studies have shown promise with spatial mode entanglement transport in specially designed custom multi-mode fibre with impressive performance \cite{Cao:2018cea, Cozzolino:2018b4b}, the distances are still orders of magnitude less than that with polarisation, and lacking the ability to integrate into existing networks, which is a crucial element of any future quantum network \cite{Lavery:2018fef}. 

% \begin{pspicture}[showgrid](0,-0.3)(3,2.3) 
%         \pnode(0,1){A}\pnode(3,1){B} 
%         \lens[n=2, lensradius=2 2, lensheight=1.5](A)(B) 
%         \drawbeam(A){1}(B) 
%         \drawbeam[beampos=0.4, linecolor=red](A){1}(B) 
%         \end{pspicture}

Here we demonstrate the transport of multi-dimensional entangled states down conventional SMF (SMF) by exploiting hybrid entangled states. We combine polarisation qubits with high-dimensional spatial modes by entangling the spin-orbit (SO) degrees of freedom of a bi-photon pair, passing the polarisation (spin) photon down the SMF while accessing multi-dimensional OAM subspaces with the other.  We show high fidelity hybrid entanglement preservation down 205 m of SMF across two two-dimensional subspaces ($2\times2$ dimensions) in what we refer to as multi-dimensional entanglement.  We quantify the channel by means of quantum state tomography, Bell inequality and quantum eraser experiments.  This work suggests an alternative approach to spatial mode entanglement transport in fibre, with the telling advantage of deployment over legacy optical networks with conventional SMF. %Furthermore, we characterise our hybrid channel for possible applications in a quantum key distribution protocol and show quantum bit error (QBER) rates as low as 0.1 and attainable key rates of up to 0.56 bits per sifted photons under the consideration of collective attacks.
%%%%%%%%%%%%%%%%%%%%%%%%%%%%%%%%
\begin{figure*}[ht!]
	\centering
	\includegraphics[width=\linewidth]{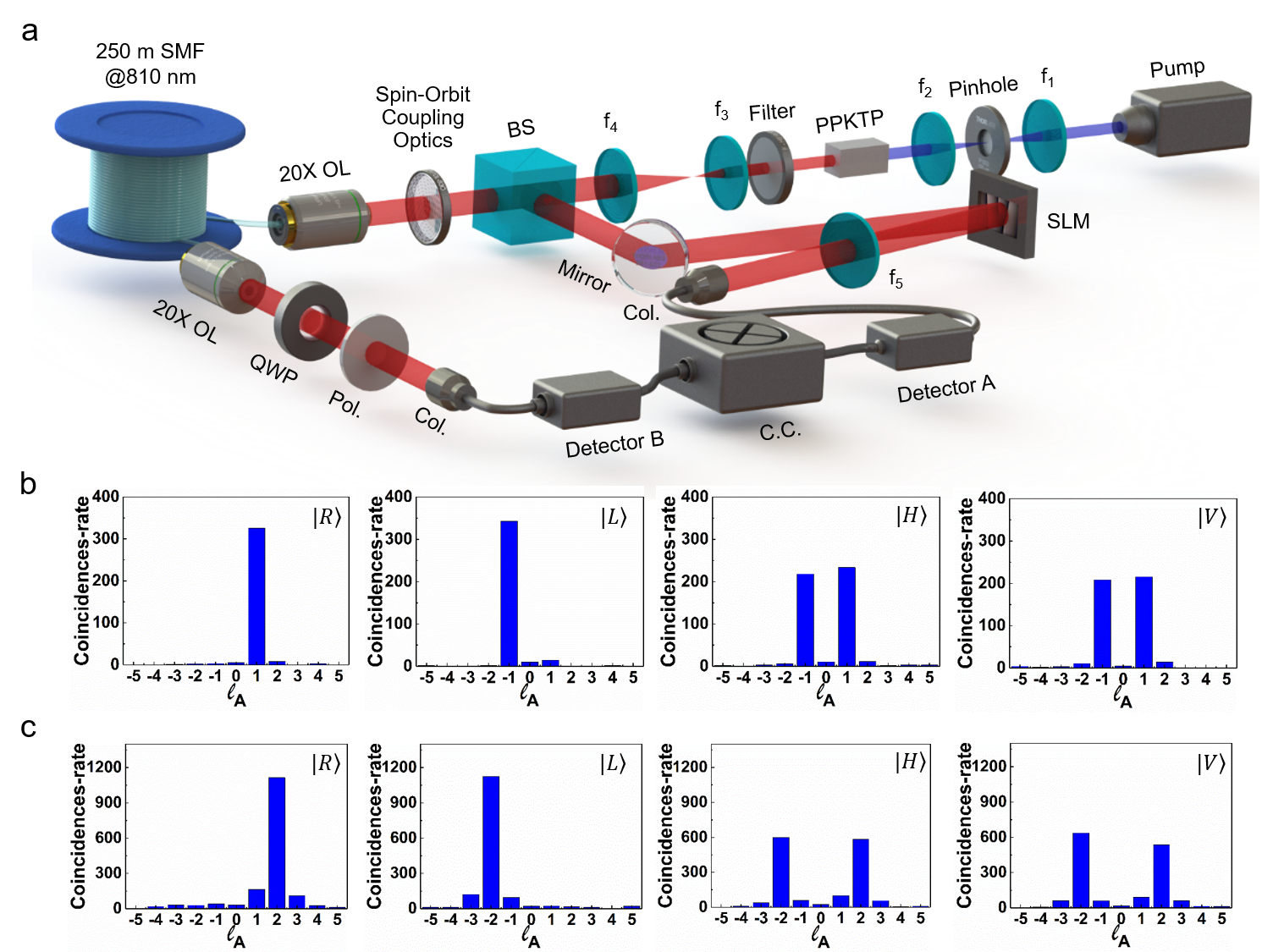}
	\caption{\textbf{Experimental setup and OAM transport.} (a) Experimental setup. Pump: $\lambda$= 405 nm (Cobolt, MLD laser diode); f: Fourier lenses of focal length $f_{1;2;3;4;5}$ = 100 mm, 100 mm, 200 mm, 750 mm, 1000 mm, respectively; PPKTP: periodically poled potassium titanyl phosphate (nonlinear crystal); Filter: band-pass filter; BS: 50:50 beam splitter; QWP: quarter-wave plate; Pol.: polarizer; SLM: spatial light modulator; Col.: collimator, f=4.51 mm; C.C.: coincidence counter. (b) The measured mode spectrum of the $\ell=\pm1$ subspace and (c) the $\ell=\pm2$ subspace after transmitting through 250 m of SMF. }
\label{setup}
\end{figure*} 
%%%%%%%%%%%%%%%%%%%%%%%%%%%%%%%%
%%%%%%%%%%%%%%%%%%%%%%%%%%%%%%%%%%%%%%%
\section{Results}
\noindent \textbf{Concept and principle.} The concept and principle of multi-dimensional spin-orbit entanglement transport through SMF is illustrated in Fig.~\ref{conceptfigure}.  Light beams carrying OAM are characterized by a helical phase front of $\exp(i\ell\theta)$ \cite{PhysRevA.45.8185}, where $\theta$ is the azimuthal angle and  $\ell \in [-\infty,\infty]$ is the topological charge. This implies that OAM modes, in principle, form a complete basis in an infinitely large Hilbert space.  However, the control of such high-dimensional states is complex, and their transport requires custom channels, e.g., specially designed custom multi-mode fibre.  On the contrary, polarisation is limited to just a two-level system but is easily transported down SMF.  Here we compromise between the two-level spin entanglement and the high-dimensional OAM entanglement to transport multi-dimensional spin-orbit hyrbid entanglement.  A consequence is that the entire high-dimensional OAM Hilbert space can be accessed, but two dimensions at a time.  We will demonstrate that in doing so we are able to transport multi-dimensional entanglement down conventional fibre.  

To see how this works, consider the generation of OAM-entangled pairs of photons by spontaneous parametric down-conversion (SPDC). The bi-photon state produced from SPDC can be expressed in the OAM basis as
\begin{eqnarray}
\ket{\psi}_{AB}=\sum_{\ell} c_{\ell}\ket{\ell}_{A}\ket{-\ell}_{B}\ket{H}_{A}\ket{H}_{B},
\label{eq:spdc}		 
\end{eqnarray}
\noindent where $|c_{\ell}|^{2}$ is the probability of finding photon $A$ and $B$ in the eigenstates $\ket{\pm\ell}$, respectively.  Subsequently one of the photons (e.g. photon A), from the N-dimensional OAM-entangled photon pair, is passed through a spin-orbit coupling optics for OAM to spin conversion, %\cite{marrucci2006optical, devlin2017spin}
 resulting in a hybrid multi-dimensional  polarisation (spin) and OAM entangled state %\cite{oe18}

\begin{equation}
	\ket{\Psi^\ell}_{AB}=\frac{1}{\sqrt{2}}\big(\ket{R}_{A}\ket{\ell_1}_{B}+\ket{L}_{A}\ket{\ell_2}_{B} \big) 
\label{eq:hybrid1}.
\end{equation} 

\noindent Here $\ket{R}$ and $\ket{L}$ are the right and left circular polarisation (spin) eigenstates on the qubit space $\mathcal{H}_{A,\text{spin}}$ of photon A and  $\ket{\ell_1}$ and $\ket{\ell_2}$ denotes the OAM eigenstates on the OAM subspace $\mathcal{H}_{B,\text{orbit}}$ of photon B. The state in Eq.~(\ref{eq:hybrid1}) represents a maximally entangled Bell state where the polarisation degree of freedom of photon A is entangled with the OAM of photon B.  Each prepared photon pair can be mapped onto a density operator $\rho_\ell = \ket{\Psi^\ell} \bra{\Psi^\ell}$ with $\ket{\Psi^\ell} \in \mathcal{H}_{A,\text{spin}}\otimes\mathcal{H}_{B,\text{orbit}}$ by actively switching between OAM modes sequentially in time: the larger Hilbert space is spanned by multiple two dimensional sub-spaces, multi-dimensional states in the quantum channel.  For simplicity we will consider subspaces of $\ket{\pm \ell}$ but stress that any OAM subspace is possible.  We can represent the density matrix of this system as

\begin{equation}
\rho_{AB}=\sum_{l=1}^{\infty} p_\ell \rho_\ell,
\end{equation}

\noindent where $p_\ell$ represents the probability of post-selecting the hybrid state $\rho_\ell$. The density matrix pertaining to system related to photon B is  

\begin{equation}
\rho_B=\text{Tr}_A(\rho_{AB})=\sum_{\ell=-\infty}^{\infty} \frac{p_\ell}{2}\big( \ket{\ell}_B\bra{-\ell}_B + \ket{-\ell}_B\bra{\ell}_B\big), \label{reduceStateB}
\end{equation}

\noindent where $\text{Tr}_A(\cdot)$ is the partial trace over photon A.  While photon B is in a superposition of spatial modes (but a single spin state), photon A is in a superposition of spin states but only a single fundamental spatial mode (Gaussian), i.e., $\ket{\psi}_A \propto (\ket{L} + \ket{R})\ket{0}$ and $\ket{\psi}_B \propto (\ket{\ell} + \ket{-\ell})\ket{H}$.  As a consequence, photon A can readily be transported down SMF while still maintaining the spatial mode entanglement with photon B.  Importantly, we stress that we use the term ``multi-dimensional'' as a proxy for multi-OAM states due to the variability of OAM modes in the reduced state of photon B: any one of these infinite possibilities can be accessed by suitable spin-orbit coupling optics, offering distinct advantages over only one two-dimensional subspace as is the case with polarisation. 

%%%%%%%%%%%%%%%%%%
\begin{figure*}[ht!]
\centering
\includegraphics[width=\linewidth]{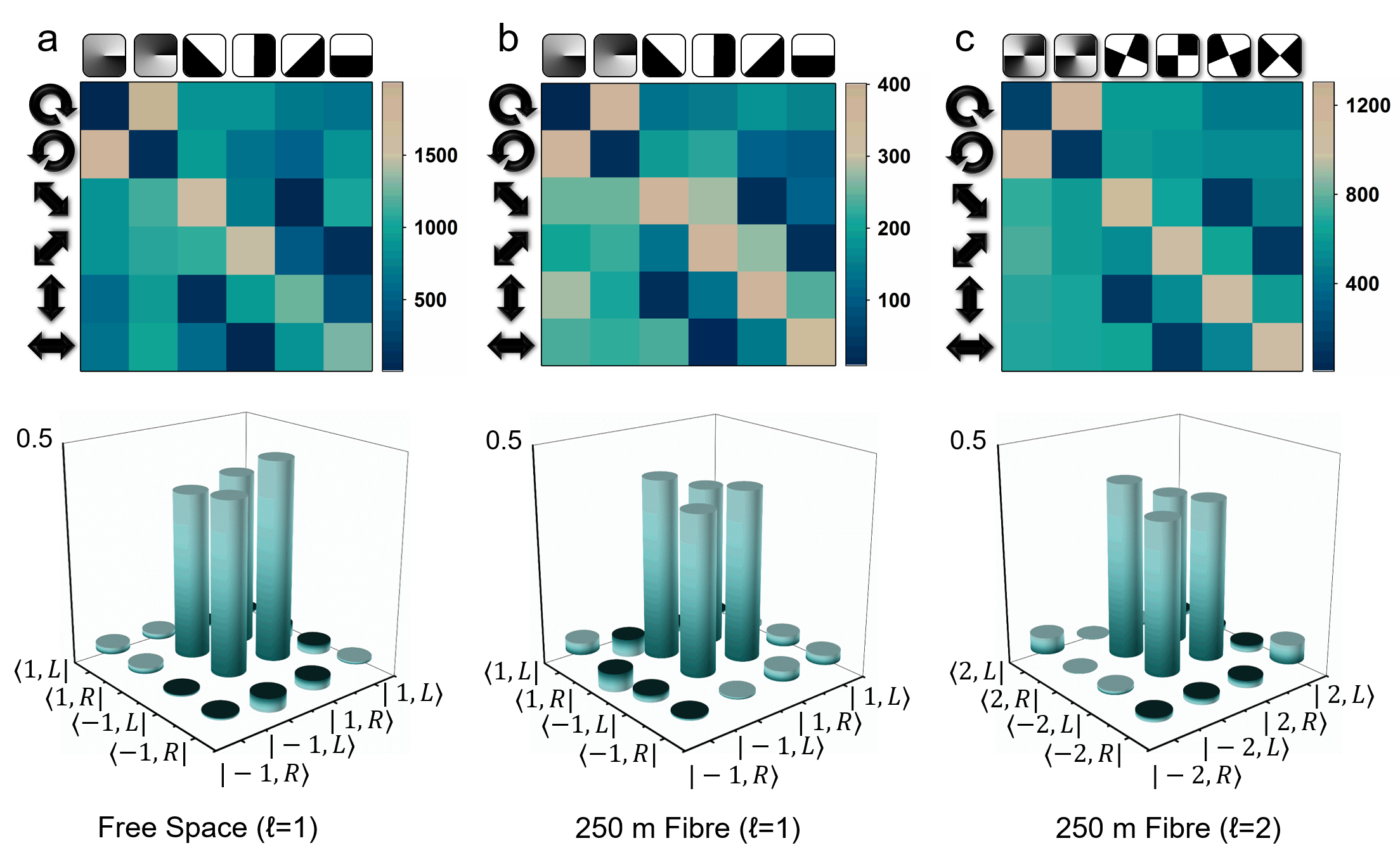}
\caption{\textbf{Quantum state tomography of hybrid multi-dimensional states.} (a) $\ell=\pm1$ in free space for system verification with (b) $\ell=\pm1$ and (c) $\ell=\pm2$ subspaces after transmitting the hybrid state through a one-sided 250 m SMF channel.  The top panels show the raw tomography data for the $6\times6$ projections, the rows representing holographic measurements on photon A and the columns representing polarisation measurements on photon B.  The bottom panels show the outcome of the tomography, a density matrix for each subspace.}
\label{tomofigure}
\end{figure*}
%%%%%%%%%%%%%%%%%%

\vspace{0.5cm}
\noindent \textbf{Implementation.}  We prepare the state in Eq.~(\ref{eq:hybrid1}) via post-selection from the high-dimensional SPDC state.  In this paper the spin-orbit coupling optics is based on q-plates \cite{PhysRevLett.96.163905}. The q-plate couples the spin and OAM degrees of freedom following

\begin{eqnarray}
\ket{\ell} \ket{R} &\xrightarrow{\text{q-plate}}& \ket{\ell - 2q}\ket{ L} \nonumber\\
\ket{\ell} \ket{L} &\xrightarrow{\text{q-plate}}& \ket{\ell + 2q}\ket{ R },\label{eq:Qplate2}
\end{eqnarray}

\noindent where $q$ is the charge of the $q$-plate. Accordingly, the circular polarisation eigenstates are inverted and an OAM variation of $\pm2q$ is imparted on the photon depending on the handedness of the input circular polarisation (spin) state.  Transmission of photon A through the SMF together with a detection acts as a post-selection, resulting in the desired hybrid state (see SI).  
%Notably, we access alternative OAM sub-spaces by concatenating multiple q-plates with half-wave retarders sand-witched between them (see SI). The resulting state becomes 
%\begin{equation}
%\ket{\Psi^{\ell+m}}_{AB}=\frac{1}{\sqrt{2}}\big(\ket{R}_{A}\ket{\ell + m}_{B}+\ket{L}_{A}\ket{-\ell-m}_{B} \big) \label{eq:hybrid2},
%\end{equation} 

%\noindent where $m$ is the total OAM charge transfer from the q-plates. In this experiment m=$\{0, 1\}$ for $\ell=1$. 
%
%The quantum state is hybrid in the sense that it is defined on a state space with each photon contributing a unique degree of freedom, i.e. $\ket{\psi}_{AB} \in \mathcal{H}_{A,spin} \otimes \mathcal{H}_{B,OAM}$ with $\mathcal{H}_{A,spin}= \textit{span}\{\ket{L}_A, \ket{R}_A\}$ and $\mathcal{H}_{B,\ell}= \textit{span}\{\ket{\ell}_B, \ket{-\ell}_B\}$, while the other degree of freedom remains separable (factorisable).
The experimental setup for multi-dimensional spin-orbit entanglement transport through SMF is shown in Fig.~\ref{setup}(a). A continuous-wave pump laser (Cobolt MLD diode laser, $\lambda$= 405 nm) was spatially filtered by a pinhole with a diameter of 100 $\mu$m to deliver 118 mW of average power in a Gaussian beam at the crystal (10-mm-long periodically poled potassium titanyl phosphate (PPKTP) nonlinear crystal), generating two lower-frequency photons by means of a type-I SPDC process. By virtue of this, the signal and idler photons had the same wavelength ( $\lambda$= 810 nm) and polarisation (horizontal). The pump beam was filtered out by a band-pass filter with the centre wavelength of 810 nm and bandwidth of 10 nm. The two correlated photons, signal and idler, were spatially separated by a 50:50 beam-splitter (BS), with the signal photon A imaged to the spatial light modulator (SLM) with lenses $f_3$ and $f_4$. After that photon A was imaged again by $f_5$ and coupled into the SMF via a fibre collimator for detection. The idler photon B, interacted with the spin-orbit coupling optics, e.g. q-plate, for orbit to spin conversion. Subsequently, the photon B was coupled into the 250 m SMF by a $20\times$ objective lens to transmit through the fibre and coupled out by another $20\times$ objective lens. The projective measurements were done by the quarter-wave plate (QWP) along with a polarizer for photon B and SLM for photon A. Photon A, encoded with multi-dimensional OAM eigenstates, was transported through free space while photon B, encoded with polarisation eigenstates was transmitted through the SMF. Finally, both photons were detected by the single photon detectors, with the output pulses synchronized with a coincidence counter (C.C.).

%%%%%%%%%%%%%%%%%%%%%%%%%%%%%%%%%%%%%%%
\vspace{0.5cm}
\textbf{Hybrid entanglement transport.}  We first evaluate the SMF quantum channel by measuring the OAM mode spectrum after transmitting through 250 m SMF.  Fig. \ref{setup}(b) and \ref{setup}(c) show the mode spectrum of the $\ell=\pm1$ and $\ell=\pm2$ subspaces, respectively.   We project photon B onto right circular polarisation, left circular polarisation, horizontal polarisation and vertical polarisation by adjusting the QWP and polarizer at the output of the fibre while measuring the OAM of photon A holographically with an SLM.  The results are in very good agreement with a channel that is impervious to OAM.  When the orthogonal polarisation states are selected on photon B ($\ket{L}$ and $\ket{R}$), an OAM state of high purity is measured for photon A, approximately 93\% for $\ell = \pm1$ and approximately 87\% for $\ell= \pm2$.  The slightly lower value for the $\ell= \pm2$ subspace is due to concatenation of two SO optics for the hybrid entanglement step (see SI and Methods).
%75\% for $\left|\ell\right|= 2$ . The latter in free space is approximately 86\% , so the purity in comparison to the free space benchmark is 87\% , indicating that the cross-talk is likely due to alignment issues and small defects in the fibre. The high purity values confirm that the hybrid OAM state is accessible even after the SMF which supports no OAM. 
%\Note{exceeding 95\% for $\ell = 1$ and 90\% for $\ell = 2$.  When superposition states are selected on photon B ($\ket{H}$ and $\ket{V}$) then superpositions of OAM are measured, as by design, exceeding 95\% for $\ell = 1$ and 90\% for $\ell = 2$.  The low levels of cross-talk are likely due to alignment issues and small defects in the fibre.}  This confirms that the hybrid OAM state is preserved and accessible even after the SMF.

To confirm the state and the entanglement, we perform a full quantum state tomography on the hybrid state to reconstruct the density matrix. Fig. \ref{tomofigure} shows the state tomography measurements and resulting density matrices for both the $\ell = \pm1$ and $\ell = \pm2$ subspaces after 250 m SMF, with the free space $\ell = \pm1$ shown as a point of comparison (see SI for more results in free space and in 2 m SMF).  The fidelity against a maximally entangled state is calculated to be 95\% for the $\ell=\pm1$ and 92\% for the $\ell=\pm2$ subspaces.  This confirms that the fibre largely maintains the fidelity of each state.   Using concurrence ($C$) as our measure of entanglement (see Methods) we find $C = 0.91$ for free space, down slightly to $C = 0.9$ for $\ell  = \pm1$ and $C = 0.88$ for $\ell = \pm2$. 
%%%%%%%%%%%%%%%%%%%%%%%%%%%%%%%%%%%
\begin{figure}[h!]
\centering
\includegraphics[width=2.8in]{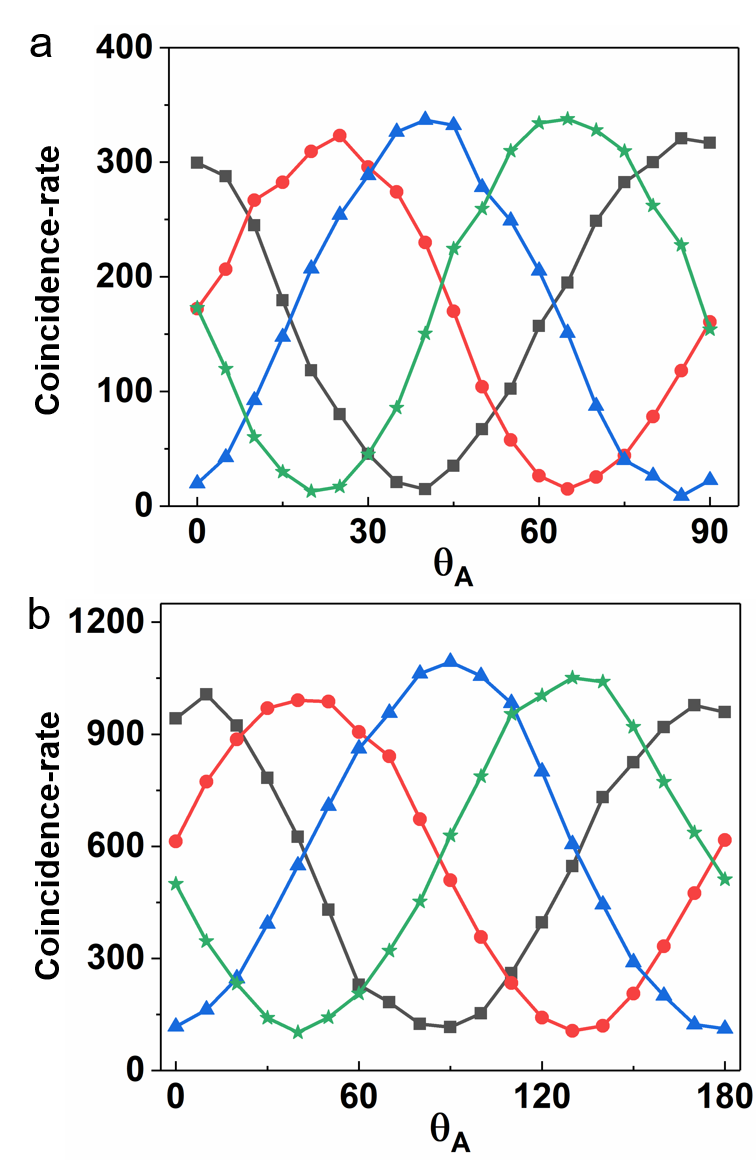}
\caption{\textbf{Hybrid state Bell violations.}  Measured correlations between photon A (polarisation) and photon B (OAM) in (a) the $\ell=\pm1$ subspace after 250 m SMF and (b) the $\ell=\pm2$ subspace after 250 m SMF.  Photon B is projected onto the states $\theta_B = \{\frac{3\pi}{2}, \pi, \frac{\pi}{2},0\}$, known to maximally violate the Bell inequality while the superposition hologram is rotated in arm A through an angle $\theta_A$.}
\label{Bellfig}
\end{figure}
%%%%%%%%%%%%%%%%%%%%%%%%%%%%%%%%%%%
%%%%%%%%%%%%%%%%%%%%%%%%%%%%%%%%%%%
\begin{figure}[h!]
\centering
\includegraphics[width=\linewidth]{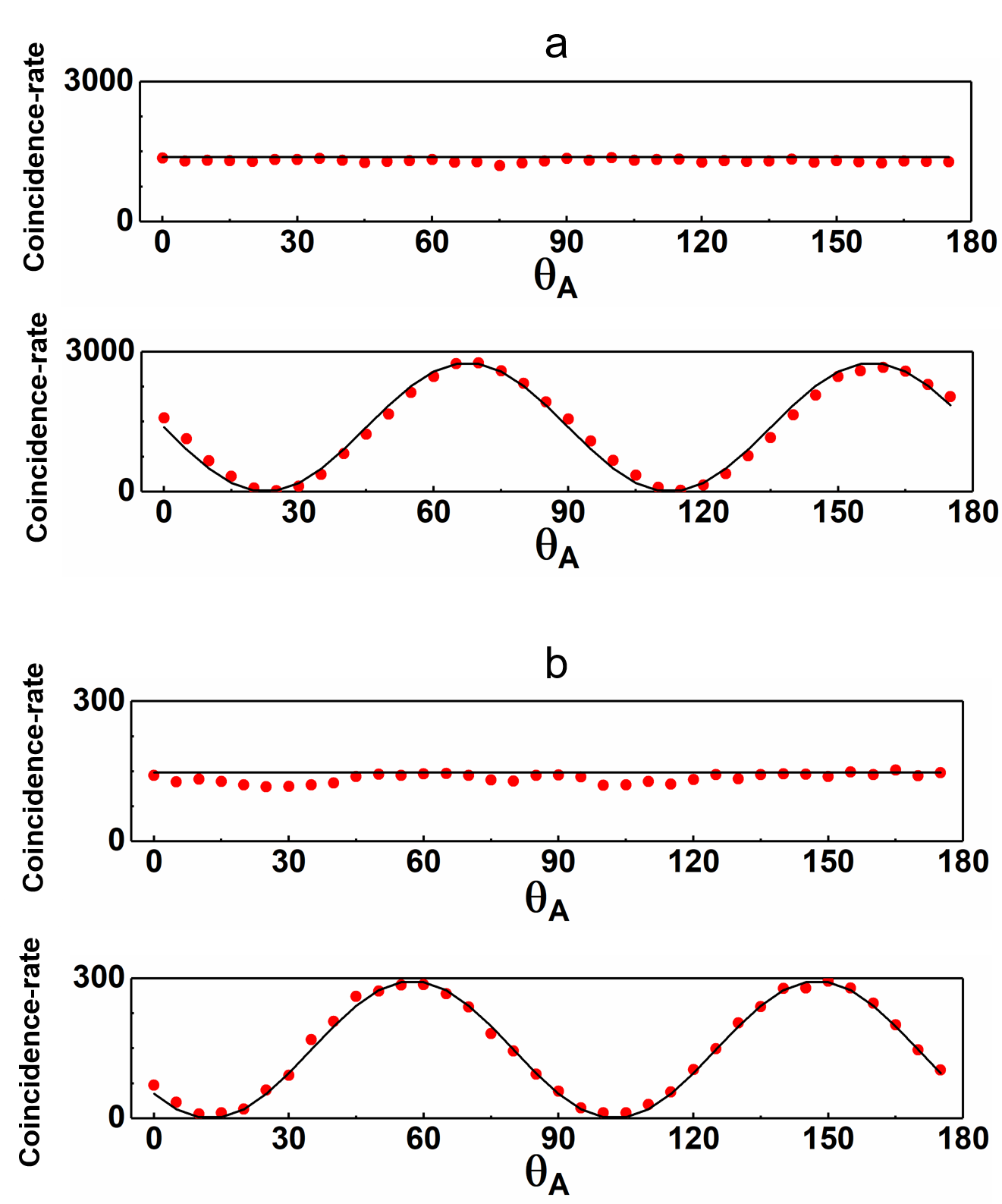}
\caption{\textbf{A hybrid quantum eraser through fibre.}  Experimental coincidence count-rates for distinguishing and erasing the OAM of photon B upon transmitting photon A through (a) free space and through (b) 250 m SMF.  OAM (path) information is introduced into the system with a QWP, and a polariser selecting one of the markers. Here we select the state $\ket{H}$ and as seen, the spatial distribution of photon B is uniform with minimal visibility. In the complimentary case, the path markers are collapsed onto a superposition $\ket{+}$ or equivilently $\ket{D}$, with the polariser and all the path information removed, manifesting as prominent visible azimuthal fringes.}
\label{eraserfig}
\end{figure}
%%%%%%%%%%%%%%%%%%%%%%%%%%%%%%%%%%%

To carry out a non-locality test in the hybrid regime we define the two sets of dichotomic observables for A and B:  the bases $a$ and $\tilde{a}$ of Alice correspond to the linear polarisation states $\{\ket{H}, \ket{V}\}$ and  $\{\ket{A}, \ket{D}\}$, respectively, while the bases $b$ and $\tilde{b}$ of Bob correspond to the OAM states $\{\cos({\frac{\pi}{8}})\ket{\ell}-\sin({\frac{\pi}{8}})\ket{-\ell}, -\sin({\frac{\pi}{8}})\ket{\ell}+\cos({\frac{\pi}{8}})\ket{-\ell}\}$ and $\{\cos({\frac{\pi}{8}})\ket{\ell}+\sin({\frac{\pi}{8}})\ket{-\ell}, \sin({\frac{\pi}{8}})\ket{\ell}-\cos({\frac{\pi}{8}})\ket{-\ell}\}$. From the data shown in Fig. 4 we calculate the CSHS Bell parameters in free space and through SMF.  We find CSHS Bell parameters in free space and in 250 m of SMF of $S=2.77\pm0.06$ and $S=2.47\pm0.09$ for the $\ell=\pm1$ subspace, respectively, reducing to $S=2.51\pm0.04$ and $S=2.25\pm0.19$ for the $\ell=\pm2$ subspace.  In all cases we violate the inequality. \\

% with the results shown in Tables \uppercase\expandafter{\romannumeral1} and \uppercase\expandafter{\romannumeral2}. In all cases we violate the inequality. \\
% As shown in Fig.~\ref{Bellfig}, we find CSHS Bell parameters of \Note{$S=(2.77\pm0.06)$, $S=(2.47\pm0.09)$, $S=(2.25\pm0.19)$ for subspace $\ell=1$ in free space, $\ell=1$ subspace and $\ell=2$ subspace after transmitting through 250 m SMF respectively.  xxxx update with all numbers in a table perhaps?  For free space and 250 m?  xxxx.}  In all cases we violate the inequality by several standard deviations.

%\begin{table}[h!]
%\begin{tabular}{|c||c|}
%\hline
%\multicolumn{2}{|c|}{$\ell=\pm1$} \\ \hline
% & S \\ \hline
%free space & 2.77 $\pm$ 0.06 \\ 
%2m & 2.71 $\pm$ 0.04 \\ 
%250m & 2.47  $\pm$ 0.09 \\ \hline
%\end{tabular}
%\caption{Measured Bell parameters for the $\ell = \pm1$ subspace.  }
%\end{table}
%
%\begin{table}[]
%\begin{tabular}{|c||c|}
%\hline
%\multicolumn{2}{|c|}{$\ell=\pm2$} \\ \hline
% & S \\ \hline
%free space & 2.51 $\pm$ 0.04 \\ 
%250m & 2.25  $\pm$ 0.19 \\ \hline
%\end{tabular}
%\caption{Measured Bell parameters for the $\ell = \pm2$ subspace.  }
%\end{table}

\vspace{0.5cm}
\noindent \textbf{A hybrid quantum eraser.}  Next we use the same experimental setup to demonstrate a hybrid quantum eraser across 250 m SMF.   We treat OAM as our ``path'' and the polarisation as the ``which path'' marker to realize a quantum eraser with our hybrid entangled photons.  We first distinguish the OAM (path) information in the system by marking the OAM eigenstates of photon B with linear polarisations of photon A. In this experiment, we achieve this by placing a QWP at $\frac{\pi}4$ before a polarisation analyser, transforming Eq. (\ref{eq:hybrid1}) to 

\begin{equation}
	\ket{\tilde{\Psi}^\ell}_{AB}=\frac{1}{\sqrt{2}}\big(\ket{H}_{A}\ket{\ell_1}_{B}+\ket{V}_{A}\ket{\ell_2}_{B} \big) 
%\lable{}
\end{equation}
% $\ket{\pm} = \frac{\ket{H}\pm\ket{V}}\sqrt{2}$
By selecting either polarisation states, $\ket{H}$ or $\ket{V}$ , the distribution of photon B collapses on one of the OAM eigenstate, $\ell_{1,2}$ , having a uniform azimuthal distribution and fringe visibility of V = 0, reminiscent of the smeared pattern that  is observed from distinguishable (non-interfering) paths in the traditional quantum eraser \cite{Walborn:2002ee5}. The OAM information can be erased by projecting photon A onto the complimentary basis of the OAM markers, i.e. $\ket{\pm} = \frac{\ket{H} \pm \ket{V}}{\sqrt{2}}$, causing the previously distinguished OAM (paths) to interfere, thus creating azimuthal fringes that can be detected with an azimuthal pattern sensitive scanner. The fringes appear with a visibility of V = 1 indicative of OAM information reduction \cite{nape2017erasing}. The appearance in azimuthal fringes of photon B is indicative of OAM information being erased from photon B. Importantly, it is noteworthy to point out that here the QWP acts as the path marker while the polariser acts as the eraser. Notably, complementarity between path information and finge visibility (V) is essential to the quantum eraser. By defining the two distinct paths using the OAM degree of freedom, we find that it is possible to distinguish ($V = 0.05 \pm 0.01$) and erase ($V = 0.98 \pm 0.002$) the OAM path information of a photon through the polarisation control of its entangled twin in free space, with only marginal lose of visibility after transmitting through 250 m SMF: we find that the entanglement is conserved with the ability to distinguish ($V = 0.11\pm 0.01$) and erase ($V = 0.93 \pm 0.01$) the OAM path information.

\section{Discussion and Conclusion}
\noindent In summary, we employed SO coupling optics in one arm of OAM entangled photons generated by SPDC and reported multi-dimensional entangled states transport down conventional SMF.  Any two-dimensional subspace of the high-dimensional space is accessible by simply changing the SO optic. In our experiment we used two sets of the SO optics, each for selecting $\ell = \pm1$, in order to reach $\ell = \pm 2$; this introduced additional distortions which reflected in the lower performance as compared to $\ell = \pm1$.  Nevertheless, even with this arrangement the entanglement was still preserved over an extended distance of 250 m, which we have demonstrated through quantum state tomography, Bell inequality violations and a novel quantum eraser experiment.  Moreover, the demonstration of two two-dimensional subspaces is double what would be possible with only polarisation entanglement.  While we used OAM states of $\{\ket{\ell}, \ket{-\ell}\}$ it is  possible to select any two orthogonal OAM states from the N-dimensional space to establish the OAM basis, i.e. $\{\ket{\ell_1}, \ket{\ell_2}\}$, ($\ket{\ell_1}\neq\ket{\ell_2}$). In addition, one can also choose any orthogonal polarisation states, for example, $\{\ket{R}, \ket{L}\}$, $\{\ket{H}, \ket{V}\}$ or  $\{\ket{A}, \ket{D}\}$. This can be done by specially designed spin-orbit coupling optics and has already been demonstrated classically \cite{Devlin896}. In this way, our work may be extended by judiciously selecting states for reducing coupling with the environment and therefore preserve the entanglement of the system over even longer distances.

%In our experiment we used two sets of the SO optics, each for selecting $\ell = \pm1$, in order to reach $\ell = \pm 2$; this introduced additional distortions which reflected in the lower performance as compared to $\ell = \pm1$.  Nevertheless, even with this arrangement the entanglement was still preserved over an extended distance of 250 m, which we have demonstrated through quantum state tomography, Bell inequality violations and a novel quantum eraser experiment.  In future experiments custom SO optics could be used Although we did not perform a quantum key distribution experiment, our setup and results make clear that this would be feasible.  allowed us to consider the performance that would be expected.  We found that quantum bit error (QBER) rates as low as 0.1 and attainable key rates of up to 0.6 bits per sifted photon.

In conclusion, we have outlined a new approach to transporting entanglement through fibre in a manner that allows deployment over a conventional network of SMF.  The result is based on hybrid entangled states, allowing access to multiple dimensions: an infinite number of two-dimensional subspaces.  Together these subspaces span the entire high-dimensional Hilbert space that would be available by spatial mode entanglement.  Our experimental demonstration over 250 m SMF and at double the dimensions available to polarisation shows that this scheme is a viable approach to circumvent the technological hurdles of deploying spatial mode entanglement.

\vspace{0.5cm}
\section*{Materials and correspondence}
Correspondence and requests for materials should be addressed to JW.

%\section*{Acknowledgments}
%xxxx.

\section*{Authors' contribution}
JL, IN, QW and AV performed the experiments, all authors contributed to data analysis and writing of the manuscript.  AF conceived of the idea.  AF and JW supervised the project.

\section*{Competing financial interests}
The authors declare no financial interests.

\newpage

\section*{Methods}
\textbf{Fidelity.}  We calculate the fidelity of our states from \cite{jozsa1994fidelity}

 \begin{equation}
 F=\mathrm{Tr}\Big(\sqrt{\sqrt{\rho_T}\rho_P\sqrt{\rho_T}} \Big)^2, \label{eq:Fidelity}
 \end{equation}
\noindent where  $\rho_T$ is the density matrix representing a target state and $\rho_P$ is the predicted (or reconstructed) density matrix. taking values ranging from 0 to 1 for $\rho_T \neq \rho_P$ and $\rho_T = \rho_P$, respectively. 

\vspace{0.5cm}
\textbf{Concurrence.}  We use the concurrence as our measure of entanglement, calculated from 
\begin{equation} 
C_{\Theta}(\rho) = \textit{max}\{0,\lambda_1-\sum_{i=2}\lambda_i\},	
\end{equation}
 where $\rho$ is the density matrix of the system being studied (mixed or pure), $\lambda$ are the eigenvalues of the operator R$= \sqrt{\rho} \sqrt{ \tilde{\rho} }$ in descending order with $\tilde{\rho}=\Theta \rho^* \Theta$ and $^*$ denotes a complex conjugation. The operator $\Theta$ represents any arbitrary anti-unitary operator satisfying $\bra{\psi}\Theta\ket{\phi}$=$\bra{\phi}\Theta^{-1}\ket{\psi}$ for any state $\ket{\phi}$ and $\ket{\psi}$, if $\Theta^{-1} = \Theta^{\dagger}$. 
 
 \vspace{0.5cm}
\textbf{Density matrix on a hybrid state space.}  The density matrix of a single photon in a two-dimensional state space ($\mathcal{H}_2$), can be represented as a linear
combination of the Pauli matrices \cite{james2005measurement}
\begin{equation}
 \rho = \frac{1}{2} \big( \mathbb{I}_0 + \sum_{3}^{k=1} b_k \sigma_{k} \big)
 \end{equation}

\noindent where $\mathbb{I}_0$ is the two-dimensional identity operator, $\sigma_{k}$ are the trace-less Pauli operators with complex coefficients $b_k$.  In this work we consider the density matrix of a hybrid entangled states similar to Eq. (\ref{eq:hybrid1}). It can be expressed as

\begin{equation}
 \rho = \frac{1}{2} \big( \mathbb{I}_A\otimes\mathbb{I}_B + \sum_{3}^{m,n=1} b_{mn} \sigma_{Am} \otimes \sigma_{Bn} \big)
\end{equation}
\noindent here $\mathbb{I}_{AB}$ is the two photon identity matrix and $\sigma_{Am}$ and $\sigma_{Bn}$ are the Pauli matrices that span the two-dimensional hybrid space for polarisation and OAM respectively.

\vspace{0.5cm}
\textbf{Quantum state tomography.}  We reconstruct each hybrid state, $\rho_\ell$, via a quantum state tomography. This entails performing a series of local projections $M_{ij}=P^{i}_{A} \otimes P^{j}_{B}$ where $P^{i,j}_{A,B}$ are projections on photon A and photon B, respectively, and using the resulting measurement outcomes to reconstruct the state \cite{jack2009precise}. The detection probabilities on a system with a corresponding density matrix ($\rho$) are
\begin{equation}
p_{ij}=Tr[M_{ij}\rho M_{ij}^\dagger].
\end{equation}
The overall projections constitute an over-complete set of measurements on the two photon subspace.

In the experiment, photon A is projected onto the spin basis states $\ket{R}$ and $\ket{L}$, along with their equally weighted superpositions of linear anti-diagonal , diagonal, horizontal, vertical polarisation states, i.e $\ket{A}, \ket{D}, \ket{V}$ and $\ket{H}$, respectively.  Similarly, photon B is locally projected onto the eigenstates $\ket{\pm\ell}$ along with superpositions
\begin{equation}
\ket{\theta}=\frac{1}{\sqrt{2}} \big( \ket{\ell} + e^{i\theta}\ket{-\ell} \big),
\end{equation}
\noindent for relative phase $\theta = \frac{3\pi}{2}, \pi,  \frac{\pi}{2} , 0$.   

\vspace{0.5cm}
\textbf{CHSH Bell violation.}  To further characterise the non-local correlations in each hybrid subspace, a violation of the John Clauser, Michael Horne, Abner Shimony, and Richard Holt (CHSH) Bell inequality \cite{clauser1969proposed} with the two photon system is used. First, we measure the photon coincidence rate as a function of  $\theta_A$ (relative phase between $\ket{\ell}$ and $\ket{-\ell}$ in arm A) while photon B is projected onto the states, $\theta_B = \{\frac{3\pi}{2}, \pi, \frac{\pi}{2},0\}$, corresponding to A, V, D, H polarisation states. The variation of the number of coincidences with the angle $\theta_A$ is in agreement with expected non-classical correlations.  We define the CHSH-Bell parameter $S$ as \cite{leach2009violation}
\begin{equation}
S = E(\theta_A,\theta_B)-E(\theta_A,\theta_B')+E(\theta_A',\theta_B)+E(\theta_A',\theta_B'), \label{eq:BellParameter}
\end{equation}
\noindent with $E(\theta_A,\theta_B)$ calculated from coincidence events
\begin{align}
& E(\theta_A,\theta_B) = \frac{A(\theta_A,\theta_B)-B(\theta_A,\theta_B)}{A(\theta_A,\theta_B)+B(\theta_A,\theta_B)}.\\
& A(\theta_A, \theta_B) = C(\theta_A,\theta_B)+C\left(\theta_A+\frac{\pi}{2},\theta_B+\frac{\pi}{2}\right),\nonumber\\
& B(\theta_A, \theta_B) = C\left(\theta_A+\frac{\pi}{2},\theta_B\right)+C\left(\theta_A,\theta_B+\frac{\pi}{2}\right).\nonumber
\label{eq:BellParameterE}
\end{align}
\noindent Here $C(\theta_A, \theta_B)$ represents measured coincidence counts. The Bell parameter can be characterized as $S\leq 2$ for separable states and $ 2<S\leq\ 2\sqrt{2} $ for maximally entangled states.

\newpage

\bibliography{ref}

\clearpage
\newpage

%%%%%%%%%%%%%%%%%%%%%%%%%%%%%%%%%%%%%%%%%%%%%%%%%
\section*{Supplementary information}
\beginsupplement{
\subsection{Additional results of mode spectrum for   subspace in free space and    subspace after transmitting through 2 m SMF}

We also evaluate the mode spectrum after transmitting through 2 m SMF ($\ell=1$ subspace) and in free space ($\ell=2$ subspace). Fig. \ref{S1figure}(a) demonstrates the mode spectrum of $\ell=1$ subspace after transmitting through 2 m fibre. The measured mode spectrum of $\ell=2$ subspace in free space is illustrated in Fig. \ref{S1figure}(b). We get $\ell=2$ subspace by using two q-plates with  inserted one half-wave plate (HWP).

%%%%%%%%%%%%%%%%%%%%%%

\begin{figure}[h]
\centering
\includegraphics[width=\linewidth]{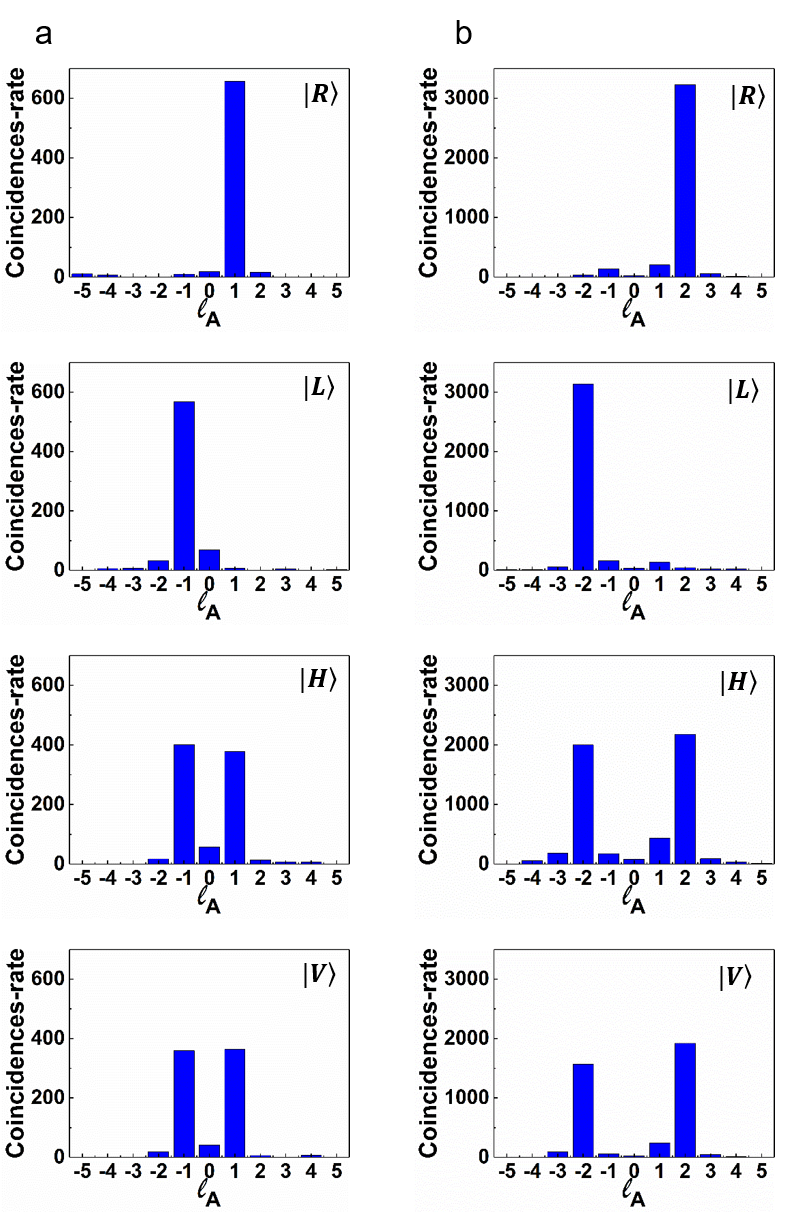}
\caption{The measured mode spectrum of (a) $\ell=1$ subspace after transmitting through 2 m SMF and (b) $\ell=2$ subspace in free space by projecting photon B onto a polarisation state and decomposing the OAM of photon A while measuring the coincidence count rate. }\label{S1figure}
\end{figure}
%%%%%%%%%%%%%%%%%%%%%%%%%

\subsection{Additional results of tomography measurements and reconstructed density matrices}

We also perform the quantum state tomographies for $\ell=1$ subspace through 2 m fibre transmission and $\ell=2$ subspace in free space as shown in Figs. \ref{S2figure}(a) and \ref{S2figure}(b), respectively. The reconstructed density matrices are demonstrated with fidelity 94\% for $\ell=1$ subspace through 2 m fibre transmission and 93\% for $\ell=2$ subspace in free space.

%%%%%%%%%%%%%%%%%%%%%%

\begin{figure}[h]
\centering
\includegraphics[width=\linewidth]{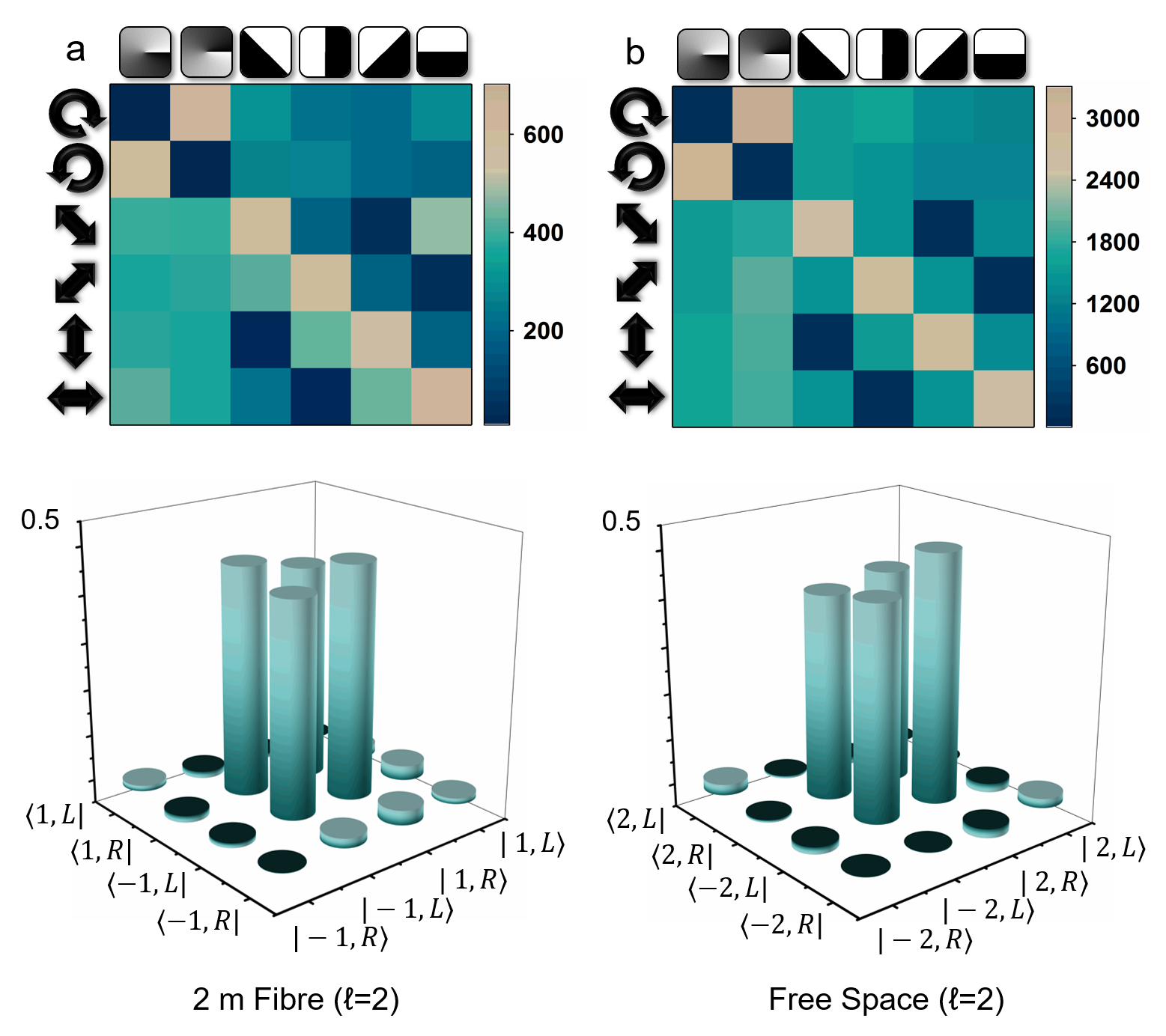}
\caption{Experimental tomography measurements upon transmitting photon A through (a) 2 m SMF for $\ell=1$ subspace and in (b) free space for $\ell=2$ subspace. The bottom panels show the reconstructed density matrices through 2 m SMF for $\ell=1$ subspace and in free space for $\ell=2$ subspace, respectively. }\label{S2figure}
\end{figure}
%%%%%%%%%%%%%%%%%%%%%%%%%
\subsection{Additional results of correlations between photon A (polarisation) and B (OAM)}

We also carry out a non-locality test in the 2 m SMF. As shown in Figs. \ref{S3figure}(a) and \ref{S3figure}(b), we get the $S=(2.71\pm0.04)$ and $S=(2.51\pm0.04)$ for subspace $\ell=1$ in 2 m SMF and $\ell=2$ subspace in free space, respectively. Furthermore, we realize quantum eraser with polarisation-OAM hybrid entangled photon in the 2 m SMF. By defining the two distinct paths using the OAM degree of freedom, we have shown that through polarisation-OAM hybrid entanglement, it is possible to distinguish ($Visibility=0.09\pm0.01$) and erase ($Visibility=0.97\pm0.002$) the OAM path information of a photon through the polarisation control of its entangled twin in 2 m SMF. 

%%%%%%%%%%%%%%%%%%%%%%

\begin{figure}[ht!]
\centering
\includegraphics[width=\linewidth]{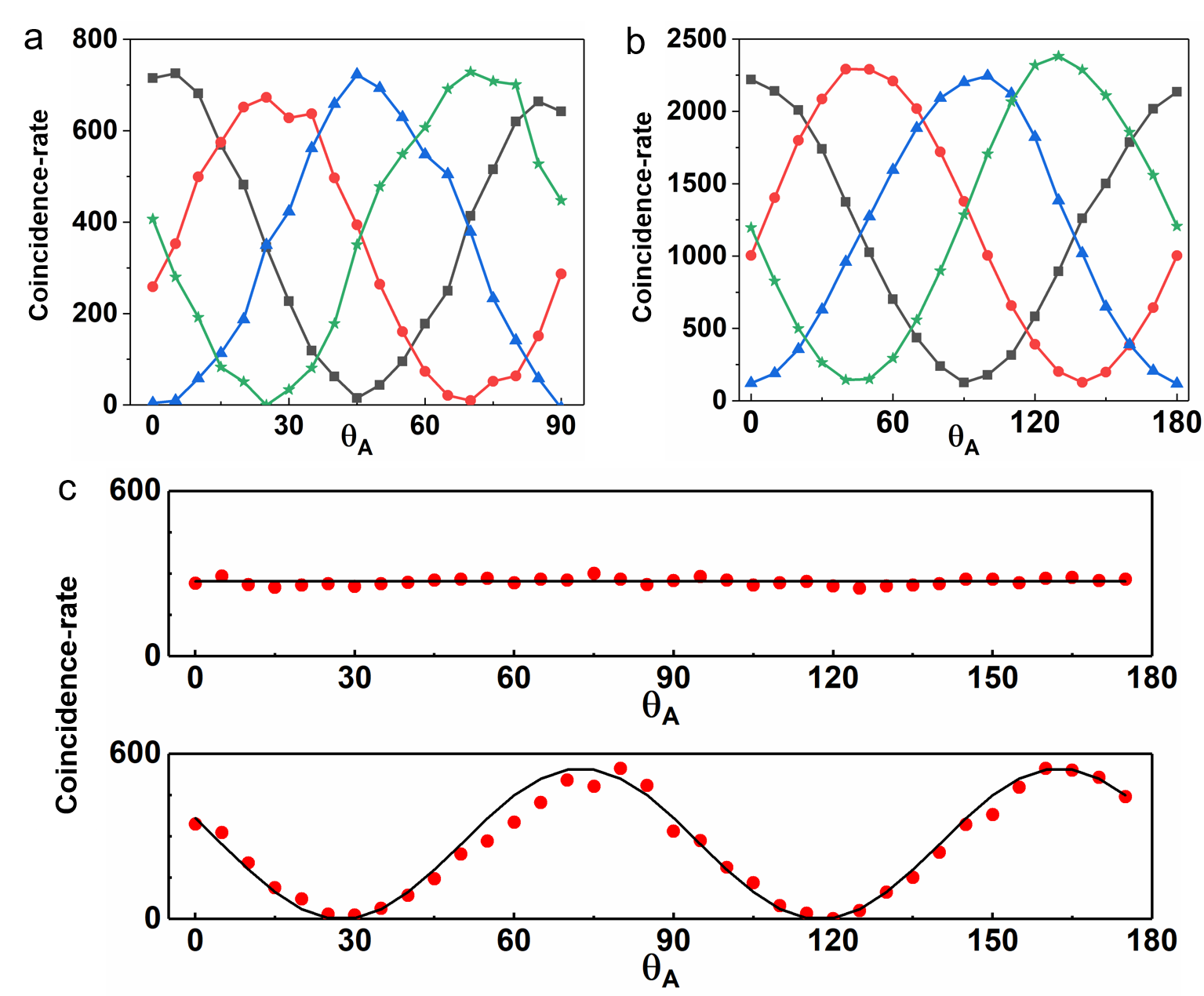}
\caption{Measured correlations between photon A (polarisation) and photon B (OAM) through (a) 2 m SMF for $\ell=1$ subspace and in (b) free space for $\ell=2$ subspace.  (c) Experimental coincidence count-rates for distinguishing and erasing the OAM of photon B upon transmitting photon A through 2 m SMF. }\label{S3figure}
\end{figure}

\subsection{Additional results of fidelity and concurrence from quantum tomography}

\begin{table}[!h]
\begin{tabular}{|c||c|c|c|c||c|c|c|c|}
\hline
           & \multicolumn{4}{c||}{$\ell$=1} & \multicolumn{4}{c|}{$\ell$=2}                                             \\ \cline{2-9} 
           & \textit{F}     & $F_n$      & \textit{C}     & $C_n$    & \textit{F}          & $F_n$         & \textit{C}          & $C_n$                \\ \hline
Free-space & 95\%  & 100\%  & 0.91  & 1.00 & 93\%             & 100\%            & 0.88             & 1.00             \\
2m         & 94\%  & 99\%   & 0.89  & 0.98 & \textbackslash{} & \textbackslash{} & \textbackslash{} & \textbackslash{} \\
250m       & 90\%  & 95\%   & 0.82  & 0.90 & 86\%             & 92\%             & 0.77             & 0.88             \\ \hline
\end{tabular}
\caption{Fidelity and concurrence values in free space, through 2m SMF and 250 SMF for $\ell=1$ and $\ell=2$ subspaces.}
\label{tableS1}
\end{table}

Here we represent all the fidelity and concurrence values in Table \ref{tableS1}. \textit{F} and \textit{C} stand for fidelity and concurrence, while $F_n$ and $C_n$ stand for Fidelity and concurrence normailized to the values in free space for corresponding $\ell$ subspaces. Since we focus on the defference between free space and SMF transmission, comparing fidelity and concurrence values normalized to free space let the actual performance in SMF stand out. If the free space fidelity is $F_1$ and the fidelity through 250 m SMF is $F_2$ then the actual performance can be described as $F_n = \frac{F_2}{F_1}$. This works for all fidelity and concurrence as well. 

%%%%%%%%%%%%%%%%%%%%%%%%%
% such as $\frac{1}{\sqrt{2}}\big(\ket{R}_{A}\ket{\ell}_{B}+\ket{L}_{A}\ket{-\ell}_{B} \big)$.

% % Hence a non-locality test was carried out. Each of two partners, A (Alice) and B (Bob) measures a dichotomic observable among two possible ones, i.e. Alice randomly measures either $\theta_{A}$ or $\theta^'_{A}$  while Bob measures $\theta_{B}$ or $\theta^'_{B}$, where the outcomes of each measurement are either +1 or −1. For any couple of measured observables ($A = \left\{ {theta,a'} \right\},B = \left\{ {b,b'} \right\}$), we define the following correlation function
% \begin{equation}
% E\left( {A,B} \right) = \frac{{N\left( { + , + } \right) + N\left( { - , - } \right) - N\left( { + , - } \right) - N\left( { - , + } \right)}}{{N\left( { + , + } \right) + N\left( { - , - } \right) + N\left( { + , - } \right) + N\left( { - , + } \right)}},
% \end{equation}
% where $N\left( { i , j } \right)$ stands for the number of events in which the observables A and B have been found equal to the dichotomic outcomes $i$ and $j$. Finally we define the parameter $S$ which takes into account the correlations for the different observables
% \begin{equation}
% S = E\left( {a,b} \right) + E\left( {a',b} \right) + E\left( {a,b'} \right) - E\left( {a',b'} \right),
% \end{equation}
% For a local realistic theory, the relation is $\left| S \right| \le S_{CHSH}  = 2$.
}
\end{document}